\documentclass[useAMS,usenatbib]{mn2e}
\usepackage{times}
\usepackage{graphicx, latexsym, amssymb, amscd, psfrag}
\usepackage{epsfig}
\usepackage{color}

\title[Dwarf galaxies in the Coma supercluster]{The evolution of dwarf
 galaxies in the Coma supercluster}
\author[Mahajan, Haines and Raychaudhury]{Smriti Mahajan\thanks{Current address: Harvard-Smithsonian
 Center for Astrophysics, MS-66, 60 Garden Street, Cambridge, 02138, MA  
 e-mail: sm@star.sr.bham.ac.uk}, Chris P. Haines, Somak Raychaudhury\\  
 School of Physics and Astronomy, University of Birmingham, Birmingham B15~2TT, UK}

\def\eg{{e.g.}}

\begin{document}

\date{}

\pagerange{\pageref{firstpage}--\pageref{lastpage}} \pubyear{2010}
\maketitle

\label{firstpage}

 \begin{abstract}
 
   We employ spectroscopic and photometric data from SDSS DR7, in a
   $500$ square degree region, to understand the evolution of
   dwarf ($\sim\!M^*\!+\!2\!<\!M_z\!<\!M^*\!+\!4$) galaxies in the Coma supercluster
   ($\rm z\!=\!0.023$). We show that in the Coma supercluster, the red dwarf galaxies
   are mostly concentrated in the dense cores of the Coma and Abell~1367 clusters, and in the
   galaxy groups embedded in the filament connecting them.  The post-starburst (k+A) dwarfs
   however are found in the infall regions of the Coma and Abell~1367
   clusters, and occasionally in galaxy groups embedded along the filament, suggesting that strong
   velocity fields prevalent in the vicinity of deep potential wells may be closely related to the
   mechanism(s) leading to the post-starburst phase in dwarf galaxies. Moreover, the blue colour of
   some k+A
   dwarfs in the Coma cluster, found within its virial radius, suggests that the star formation
   in these galaxies was quenched very rapidly in the last $500$\,Myr. 
   More than 60\% of all red dwarf galaxies in the supercluster have 0--3\AA~of
   H$_\delta$ in absorption, which suggests that a major episode of star formation occurred in
   a non-negligible fraction of these galaxies, ending within the last Gyr, allowing them
   to move to the red sequence. The distribution of the blue dwarf galaxies in the Coma
   supercluster is bimodal in the EW(H$_\alpha$)--EW(H$_\delta$) plane, with one population
   having very high emission in H$_\alpha$, and some emission in H$_\delta$. A sub-population
   of blue dwarfs is coincident with the red dwarfs in the EW(H$_\alpha$)--EW(H$_\delta$) plane,
   showing absorption in H$_\delta$ and relatively lower emission in H$_\alpha$. 
   We suggest that a large fraction of the latter population are the progenitors of the passive
   dwarf galaxies that are abundantly found in the cores of low-redshift rich clusters such as
   Coma.
          
 \end{abstract}


 
 \section{Introduction}
 
 In hierarchical models of structure formation, dwarf galaxies are the
 `building blocks' of the more massive galaxies. Moreover, the
 vulnerable nature of these galaxies, which are most abundant in the
 Universe, makes them an excellent probe for exploring the impact of
 the environment on evolution of galaxies. A reliable census of
 dwarfs, and detailed knowledge of their properties is required, over
 a wide range of epochs, to understand the coeval evolution of the
 large-scale structure and the galaxies therein.
 
 However, unlike their giant counterparts, dwarfs have received little attention in the
 literature, possibly due to the challenges encountered in observing these faint and low surface
 brightness galaxies. Nevertheless, studies conducted in the last decade or so have been able to
 exploit the multiplex gain of wide-field multi-fibre spectrographs \citep[\eg][]{pogg01},
 while others have targeted the low-luminosity \citep[\eg][and references therein]{lisker07,smith08}
 and low surface brightness \citep[\eg][]{chiboucas10} galaxy population in nearby clusters
 \citep[\eg][and references therein]{lisker07,smith08}. 
 
 Large samples of dwarf galaxies have now been exploited to show that passive dwarfs are found
 only in very dense environments, or accompanying a more massive companion
 \citep[\eg][]{haines07}. 
 Studies of HI gas content in dwarf galaxies do not find any correlation between the HI gas surface
 density and star formation rate \citep{roychaudhury09} or star formation efficiency
 \citep{bothwell09}, suggesting that dwarf galaxies are star-forming everywhere. Using HI surface
 density normalized by stellar mass to quantify star formation efficiency, \citet{bothwell09}
 suggest that mass dependent quenching mechanism(s) are likely to play a dominant role in galaxy
 evolution.
 These results imply that transformations in the observable properties of dwarf galaxies may
 occur only in specific environments, and hence galaxy populations in the process of transformation
 should be confined to such regions. The post-starburst (or k+A) galaxies are among the best
 candidates for such transitional populations.
 
 The k+A galaxies show strong Balmer absorption but lack emission lines in their spectra. 
 Such galaxies are mostly found in and around galaxy clusters at $\rm z\!\sim\!0$
 \citep[\eg][]{mahajan09b}
 and at higher redshifts \citep[\eg][but see \citet{goto05,yan09} for an alternative
 view]{dressler82,pogg04}. In the Coma and Abell~2199 supercluster ($\rm z\!=\!0.023$),
 dwarf galaxies show a stronger relation between star formation rate and local galaxy density
 than is found in the more massive galaxies \citep{haines06,mahajan09b}, while elsewhere 
 \citep{barazza09} a strong colour-density relation is found in Abell 901/902 system
 ($\rm z\!=\!0.165$). Clusters and rich groups at intermediate and high redshifts ($\rm z\!\sim\!0.5$)
 have more luminous ($M_V\!\leq\!-20$) k+A galaxies, which seem to be missing in rich clusters at
 $\rm z\!\sim\!0$ \citep{zabludoff96,pogg04,nolan07}.   
         
 In \citet{mahajan09b}, we combined the optical photometric and
 spectroscopic data (from the Sloan Digital Sky survey data
 release 7, SDSS DR7) with the 24\,$\mu$m MIPS observations to study
 the star-formation and active galactic nuclei (AGN) activity in
 galaxies across one of the nearest ($\sim\!100$ $h_{70}^{-1}$ Mpc)
 supercluster, namely Coma.  In this article we use a subset of data from SDSS DR7 to
 understand the origin of dwarf k+A galaxies, and their evolutionary link to the passive red dwarf
 galaxies that are observed in abundance in the Coma cluster \citep{jenkins07}. We briefly
 describe the dataset and definition of different galaxy populations in the following section. In
 \S\ref{colour} and \S\ref{evolution} we present our analysis, finally discussing the implications
 of our analysis in \S\ref{discussion}. 
 
 We adopt the
 concordance cosmology ($\Omega_{\Lambda}\!=\!0.70; \Omega_{M}\!=\!0.30; h\!=\!0.7$) for calculating
 the magnitudes and distances. We note that at the redshift of Coma
 (z\,$=\!0.023$), our results are independent of the choice of cosmology. Different
 techniques give varying estimates for the virial radius of the Coma cluster such that
 $2\!<\!R_{\rm virial}\!<\!3\,h_{70}^{-1}$Mpc. For this paper the virial radius for the Coma
 cluster is assumed to be $2$\,$h_{70}^{-1}$Mpc based on the weak-lensing analysis of
 \citet{kubo07}. Any reference to the cluster core in this article imply region
 $\lesssim\!0.2\!R_{\rm virial}$ of the cluster centre. Throughout this article, we will refer to
 the pair of clusters Coma and Abell~1367, along with the associated filament of galaxies, as the
 Coma Supercluster.      

 \section{Data}
 \label{data}
 
 The optical photometric and spectroscopic data acquired by the SDSS DR7 \citep{abazajian09} are
 used to select the galaxies belonging to the Coma supercluster from the SDSS spectroscopic
 galaxy catalogue only, requiring the member galaxies to be within $170\!\le$RA$\le\!200$
 deg and $17\!\le$Dec$\le\!33$ deg on the sky, and with a radial velocity within 3,000
 km~s$^{-1}$ of the mean redshift of the Coma supercluster. 
 All our galaxies are brighter than SDSS magnitude $r\!=\!17.77$ ($\sim$M$^{*}\!+\!4.7$ for
 Coma cluster), which is the magnitude limit of the SDSS spectroscopic catalogue. In order to
 select the dwarf population by stellar mass, we define galaxies with $z\!\geq\!15 (M^*\!+\!2)$
 as `dwarfs' throughout this work since it is relatively extinction free comapared to other SDSS bands.
 However, we note that for the sample presented here, the $r$ and $z$ bands are linearly correlated.
 This gives us a sample of 3,050 dwarf galaxies which is used in 
 this article.
 
 The k+A galaxies, defined to have $>\!3$\,\AA~of H$_\delta$ in absorption and $<\!2$\,\AA~of
 H$_\alpha$ in emission, are taken from the catalogue compiled in \citet{mahajan09b}. Hereafter,
 we consider all galaxies with EW(H$_\alpha$)\,$>\!2$\,\AA~as star-forming, and any reference to
 galaxy colour in context of this work refers to the broadband ($g\!-\!r$) colour (see \S\ref{colour}). 
 
 \section{Colour and environment}
 \label{colour}
 
 \begin{figure}
 \centering{
 {\rotatebox{270}{\epsfig{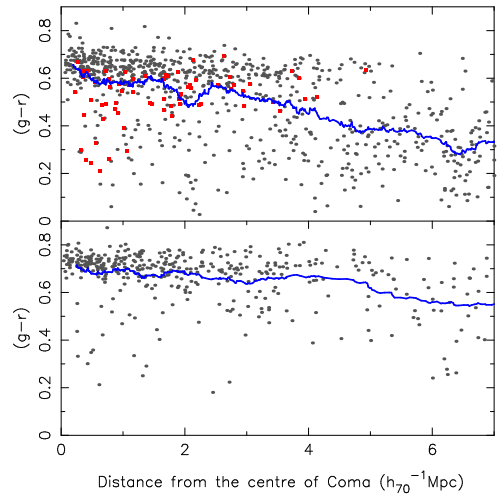}}}}
 \caption{The radial dependence of $(g\!-\!r)$ colour of {\it{(top)}} dwarf ({\it{grey points}};
 $z\!>\!15$) and k+A ({\it{red squares}}),
 and {\it{(bottom)}} giant ($z\!\leq\!15$) galaxies within 7 $h_{70}^{-1}$ Mpc of
 the centre of the Coma cluster. The solid line shows the running mean with 50 galaxies per bin.
 The red sequence for just the dwarfs in Coma cluster extends out to the virial radius
 ($2\,h_{70}^{-1}$\,Mpc)-- same as for the more massive galaxies. Most of the blue k+A dwarfs
 lie within R$_{\rm virial}$. } 
 \label{cmr}
 \end{figure}
 
  \begin{figure*}
 \centering{
 {\rotatebox{270}{\epsfig{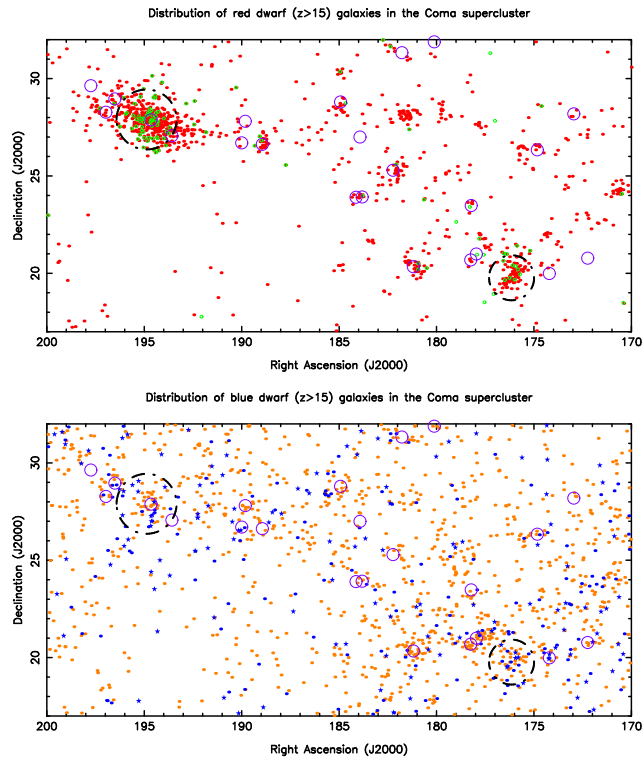}}}}
 \caption{{\it{(top)}} The distribution of red ({\it{red points}}) and k+A
 ({\it{green points}}) dwarf ($r\!\leq\!17.77; z\!>\!15$) galaxies in the Coma supercluster
 ($cz\,=\,6,973\!\pm\!3,000$ kms$^{-1}$). The {\it{purple circles}}
  represent galaxy
  groups from literature, that lie in the same redshift slice, and the dashed circles are centered
  on cluster centres with a radius of 3\,R$_{\rm virial}$. {\it{(bottom)}} The same for blue galaxies
  with EW(H$_\alpha$)\, $\!<\!50$\AA~\& EW(H$_\delta$)\,$\!\leq\!0$\AA~{\it{(orange points)}},
  strabursts ($50\!\leq\,$EW(H$_\alpha$)\,$\!\leq\!80$\AA; {\it{blue points}}) and {\em{extreme}}
  starbursts (EW(H$_\alpha$)\,$\!>\!80$\AA~{\it{(blue stars)}}; also see Fig.~\ref{ha-hd}).
  While the blue dwarfs trace the entire supercluster, the red ones are abundant in the high density
  regions only. The k+A dwarfs are mostly confined to the vicinity of the two clusters, and
  occasionally the galaxy groups. }
 \label{scl}
 \end{figure*}
 In Fig.~\ref{cmr}, we plot the radial dependence of $(g\!-\!r)$ for all dwarf ($r\!\leq\!17.77; 
 z\!>\!15$) galaxies found in and around the Coma cluster. The mean colour of dwarfs changes
 from 0.3 around 3R$_{\rm virial}$ away from the centre, to 0.7 in the core of the cluster. It is
 also interesting to note that even with dwarfs alone, the red sequence for the Coma cluster
 extends beyond the virial radius, suggesting that the potential well of the cluster effectively
 contributes to the evolution of all galaxies as far as the cluster periphery. 
 
 The k+A galaxies
 in the vicinity of the Coma cluster span a wide range in $(g\!-\!r)$ colour, such that most of the
 blue k+As lie within R$_{\rm virial}$. The blue colour of some k+As suggests that the star
 formation in these galaxies has been very rapidly quenched no longer than $500$\,Myr ago
 \citep[also see][]{pogg04}. In contrast, a Gyr should have
 elapsed since the termination of the last major episode of star formation in the red k+As,
 allowing for their observed $(g\!-\!r)$ colour. The k+A dwarfs
 show a relatively scattered distribution compared to red dwarfs. Assuming that some of these galaxies are
 seen in projection along the line of sight, it is fair to state that the k+As prefer
 intermediate density regions found in the outer regions and vicinity of rich clusters and groups
 \citep[Figs.~\ref{cmr} \& \ref{scl}; also see fig.~12 of][for positions of k+A galaxies in the
 clusters]{mahajan09b}, indicating a possible link between the processes contributing to the k+A
 phase in dwarf galaxies and the depth of the cluster potential.

 We use the $(g\!-\!r)\!-\!M_r$ colour--magnitude diagram to separate red and blue galaxies in the
 Coma supercluster, where, the blue galaxies are defined to have $(g\!-\!r)$ colour one
 mean absolute deviation (MAD) below the fitted red sequence. For consistency in this work we fitted
 this red sequence using dwarf galaxies only. This red sequence on an average assigns
 $(g\!-\!r)\!\sim\!0.64$\,mag to a galaxy of $M_r\!=\!-17$ mag. The location of the red and blue
 dwarf galaxies across the Coma supercluster is shown in Fig.~\ref{scl}. The k+A galaxies and galaxy groups
 found in the NASA Extra-galactic database (NED) \citep[see][for the selection criteria]{mahajan09b}
 are shown explicitly. The red dwarf galaxies trace the rich structures-- clusters,
 groups and the filament,
 suggesting a stronger star formation-density relation than their blue counterparts.

 \section{Spectral diagnostics} 
 \label{evolution}
 
 In the absence of dust, the equivalent width (EW) of H$_\alpha$ is a good indicator of the
 current star formation rate (SFR) of galaxies. On the other hand, intrinsic H$_\delta$ absorption
 is a characteristic feature of A-type stars, which have a lifetime of 1--1.5 Gyr. Hence, a galaxy
 that experienced a burst of star formation 1--2 Gyr ago is likely to show strong absorption in
 H$_\delta$. But star-forming HII regions in galaxies with ongoing starburst can
 produce enough ionizing radiation to fill-in the H$_\delta$ absorption line.
 
\begin{figure}
 \centering{
 {\rotatebox{270}{\epsfig{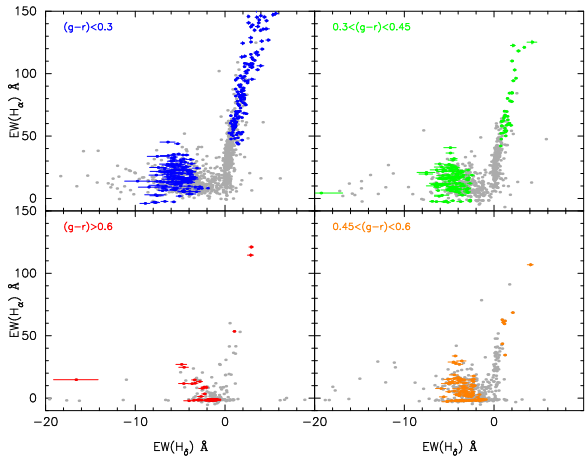}}}}
 \caption{The distribution of dwarf galaxies found in the Coma supercluster on the
 EW(H$_\alpha$)--EW(H$_\delta$) plane in four bins of $(g\!-\!r)$ colour, where colour becomes redder
 clockwise from the top left panel. The {\it{grey points}} show all galaxies belonging to the given
 panel, while the {\it{coloured points}} with error bars are galaxies with $\leq\!15\%$ measurement
 uncertainty in EW(H$_\alpha$) {\em{and}} EW(H$_\delta$). The red dwarfs, which are abundant in
 the high density regions of the supercluster (see Figs.~\ref{cmr} and \ref{scl}),
 mostly show absorption in H$_\delta$ with a few showing some H$_\alpha$ emission,
 suggesting ongoing star formation. The blue dwarfs on the other hand show at least
 two sub-populations, one of which show similar spectral features as their red counterparts.
 The other population of blue dwarf galaxies show very strong emission in H$_\alpha$ and
 sometimes also in H$_\delta$, indicating a very recent or ongoing burst of star formation.  }
 \label{ha-hd}
 \end{figure}

 In Fig.~\ref{ha-hd} we show the distribution of dwarfs in the Coma supercluster in the
 EW(H$_\alpha$)--EW(H$_\delta$) plane, segregated into four bins of $(g\!-\!r)$ colour.
 The four colour bins are such that the both the redder bins contain galaxies belonging
 to the red sequence. As galaxies become redder towards the cluster core (Figs.~\ref{cmr}
 \& \ref{scl}) the narrow ridge formed by blue galaxies having very strong H$_\alpha$
 emission but negligible H$_\delta$ absorption vanishes.
 In dealing with the spectral features it is important to check whether the two
 populations seen in this EW(H$_\alpha$)--EW(H$_\delta$) plane result from measurement
 uncertainties. In order to do that, we split the galaxies in each panel into those that
 have measurement uncertainty according to SDSS, $\leq\!15\%$ of the measured
 EW for both H$_\alpha$ and H$_\delta$ (the coloured points in Fig.~\ref{ha-hd}) and those
 that do not (grey points in Fig.~\ref{ha-hd}). As expected, as the emission features become weak
 for the passively evolving red dwarfs, more galaxies inhabit the uncertain zone. 

 However among the blue, star-forming class, there exists at least two
 sub-populations, one that of the starburst dwarfs, which line-up in a narrow ridge on the right
 hand side of the EW(H$_\alpha$)--EW(H$_\delta$) plane, and the other of normal star-forming
 and post-starburst galaxies that have significant absorption in H$_\delta$.
 The latter class of blue dwarfs overlaps with the red ones, showing relatively
 low EW(H$_\alpha$) and some absorption in H$_\delta$. Although most of these galaxies have continuous
 star formation histories with a low SFR, a sub-population might have been quenched recently
 \citep[$\lesssim\!1$ Gyr old; ][]{pogg04} as they are assembled into clusters via the large-scale structure
 \citep[\eg][]{mahajan10}, thus terminating the starburst.
 From Figs.~\ref{cmr} and \ref{scl} it is also evident that although most of the blue dwarfs
 within 3R$_{\rm virial}$ of the cluster centres have low H$_\alpha$ emission ($<\!50$\AA),
 such galaxies are found elsewhere in the supercluster as well.
 
 The red dwarfs showing H$_\delta$ absorption similar to the blue galaxies could
 be (i) dominated by metal-rich stellar populations \citep{smith08,mahajan09a}, or
 (ii) transformed large spiral galaxies, whose disks have been removed after
 they fell into the cluster \citep[\eg][]{abadi99,barazza09}, due to the impact of environmental
 mechanisms such as ram-pressure stripping \citep{gunn72}, or (iii) star-forming dwarf
 galaxies in the early phases of transformation \citep[][and references
 therein]{lin83,boselli08}. We discuss these possibilities further in \S\ref{discussion}.

 In the Coma cluster, most ($\sim\!71\%$) of the dwarf galaxies are red (Fig.~\ref{cmr})
 and show no H$_\alpha$ emission (Fig.~\ref{ha-hd}),
 while the same fraction is only $\sim\!25\%$ in the neighbouring Abell~1367 cluster. 
 Given that a significant fraction of blue galaxies with more than $3$\,\AA~of absorption in
 H$_\delta$, also have $<\!20$\,\AA~of emission
 in H$_\alpha$ (Fig.~\ref{ha-hd}), it is likely that star formation in some of them was abruptly
 quenched. 

 \section{Discussion}
 \label{discussion}
 
 In this work we attempt to understand the origin of k+A galaxies, and the possible evolutionary
 link between this transitional population and the red dwarfs found abundantly in the Coma cluster
 \citep[\eg][]{jenkins07,chiboucas10}. We use the publicly available SDSS DR7 data for the entire
 supercluster, with a view to studying the influence of large-scale structure, if any, on the evolution of 
 the k+A galaxies in the low redshift Universe. In agreement with other studies of cluster dwarfs
 in the nearby Universe \citep[\eg][]{pogg04,boselli08}, we find that the k+A galaxies
 prefer the intermediate density regions in the vicinity of rich structures. General studies of k+A
 galaxies at z\,$\sim\!0$ also reveal that they preferentially avoid very low and high density
 regions \citep{balogh05,goto05}.

 If star formation in a galaxy is truncated by some mechanism affecting the halo gas, star
 formation is quenched on long time scales of $\sim\!1$ Gyr \citep[\eg][]{kodama01,kawata08}.
 However if the star formation is suddenly quenched, for instance when a galaxy encounters the cluster
 environment and is stripped not only of the reservoir of gas but also the supply for
 ongoing star formation, the SFR declines on much shorter timescales of $\sim\!100\!-\!150$ Myr.
 Thus one is more likely to observe the transient populations, such as the blue passive galaxies
 \citep[\eg][]{mahajan09a} and k+A glaxies in the latter scenario. This is also consistent with the
 idea that an infalling galaxy can lose most of its gas in a starburst or violent encounters
 with fellow infalling galaxies on the cluster periphery \citep[][and references therein]{mahajan10}
 and subsequently evolve in a passive fashion as it settles into the cluster core, on a timescale
 equivalent to a few times the cluster crossing time ($\sim\!1$ Gyr). 
 
 For the dwarf galaxies falling into clusters and groups in the
 Coma supercluster, the velocity field varies significantly depending upon the depth of the
 potential well they are approaching. Hence, it is likely that the galaxies falling into deeper
 potentials undergo a burst of star formation in which they either exhaust all the gas content,
 and/or lose it through interactions with the intra-cluster medium (ICM), thus experiencing a sudden
 and rapid cessation in star formation \citep[\eg][also see Fig.~\ref{image}]{tzanavaris10}.
 On the other hand, galaxies falling into relatively shallower potentials are likely to suffer slow
 quenching. This explains the presence of dwarf k+A galaxies  mostly in the vicinity of the
 two rich clusters, Coma and Abell~1367, and occasionally in galaxy groups embedded in the filament
 crossing them (Fig.~\ref{scl}).

 In the Coma supercluster we find that around 86\% (78\%) of all the k+A dwarfs are found within 10
 (5) $h_{70}^{-1}$ Mpc of the centre of the clusters, suggesting a close link between the transitional
 k+A population and the cluster environment. Consistent with previous results \citep{pogg04},
 we find that most of the blue k+A dwarfs in the vicinity of the Coma cluster are found within its
 virial radius. The blue colour of some k+As suggests that the star formation in these galaxies has been
 quenched rapidly within the last $500$\,Myr.
 These observations are similar to those made for the comparable Virgo cluster.
 \citet{boselli08} studied the evolution of dwarf ($L_H\!<\!10^{9.6}$ M$_\odot$)
 galaxies in Virgo by comparing their UV to radio spectral energy distributions (SEDs) to
 multi-zone chemospectrophotometric models. Their results suggest that not only is the star
 formation in
 the dwarf galaxies entering the Virgo cluster quenched on short ($\sim\!150$ Myr) time-scales, most
 of the luminous late-type spirals might also be transformed to relatively massive dEs in a similar
 fashion.

\begin{figure}
 \centering{
 {\rotatebox{270}{\epsfig{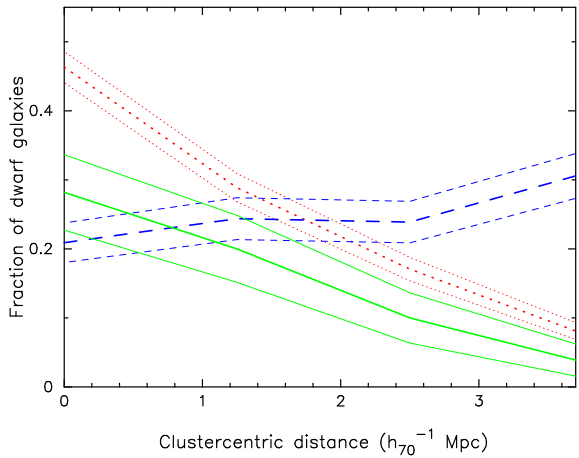}}}}
 \caption{The distribution of the red ({\it{red dotted line}}), blue
 ({\it{blue dashed line}}) and k+A ({\it{green solid line}}) galaxies as a function of
 clustercentric radius from the centre of the Coma cluster. 525, 157 and 67 galaxies
 contribute to each of the 3 curves respectively. The {\it{thin lines}} corresponding to each
 distribution represent the $\pm\!1\!\sigma$ scatter, assuming binomial statistics. All curves are
 individually normalized to unity. }

 \label{radial}
 \end{figure}

\begin{figure*}
 \centering{
 {\rotatebox{0}{\epsfig{file=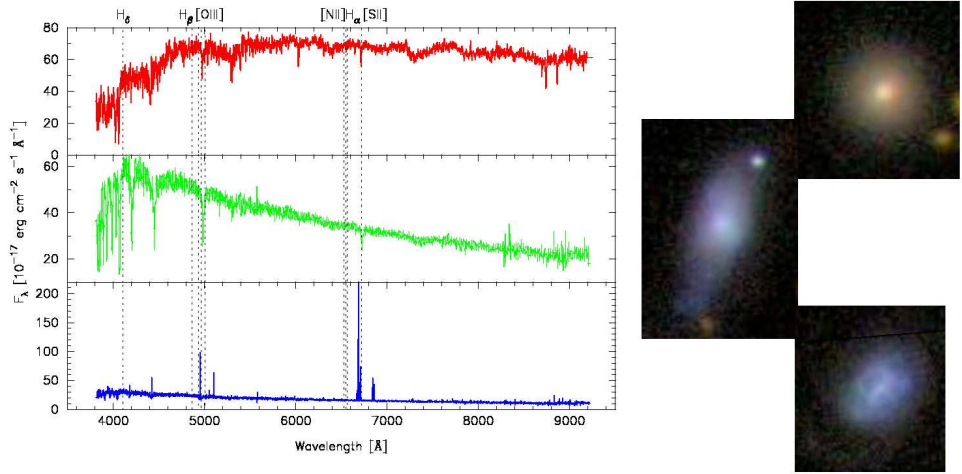,width=13cm}}}}
 \caption{Typical SDSS spectra of a red ({\it{top}}), k+A ({\it{middle}}) and
 blue star-forming ({\it{bottom}}) dwarf galaxy in Coma, and their respective images
 ({\it{right}}). All images are obtained in colour from the SDSS website. In this article we
 suggest that in the Coma supercluster, transformation of infalling
 star-forming dwarfs into passive ones occurs via a k+A phase, whereas the dwarf galaxies falling
 into relatively shallow potential wells of galaxy groups embedded elsewhere in the supercluster are
 likely to suffer slow quenching.  }
 \label{image}
 \end{figure*}

 In Fig.~\ref{radial} we find evidence to support a similar scenario in the Coma cluster.
 While the fraction of red dwarfs steadily increases from 4 $h_{70}^{-1}$
 Mpc ($2$\,R$_{\rm virial}$) to the centre of the cluster, the population of blue dwarfs
 declines steadily within the virial radius. The distribution of k+A dwarf galaxies follows the red
 dwarfs. Since
 the k+A phase is relatively short-lived \citep[$\sim\!0.5$ Gyr;][]{balogh05} in a galaxy's lifetime,
 the observation of even a small fraction of such a transitional population suggests that a much larger
 proportion of all galaxies in the region may have passed through this stage.
 
 Stellar population synthesis models suggest that galaxies with k+A spectral features (strong Balmer
 absorption and lack of emission) are best modelled as a post-starburst \citep[][and
 references therein]{dressler92,pogg99}, i.e. star formation is completely ceased following a burst.
 Studies of large-scale inter-cluster filaments provide statistical evidence to show that galaxies
 infalling into clusters are likely to undergo a starburst on the outskirts of clusters, before they are
 assimilated in their cores \citep[\eg][and references therein]{porter08,mahajan10}. These results imply
 that bursts of star formation in galaxies falling into clusters are inevitable consequences of the
 hierarchical assembly of structures. This is mainly dependent on two factors:
 (i) the gas content of the infalling galaxies, and (ii) the velocity field around the cluster
 \citep[\eg][]{mahajan10b}. 
 This scenario is further supported by the SDSS based study of k+A galaxies by \citet{balogh05},
 who find that the $(u\!-\!g), (r\!-\!K)$ colours and the H$_\delta$ EW of k+A galaxies are
 consistent with models where $>\!5\%$ of the stellar mass of galaxies is formed in a recent
 starburst. 

 We find that 7.8\% of all dwarf galaxies within 3R$_{\rm virial}$ of the Coma
 cluster are k+As, while in the relatively smaller Abell~1367 this fraction declines to 4.1\% within
 the same aperture. Furthermore, excluding the clusters, the fraction of k+A galaxies
 in the rest of the supercluster (Fig.~\ref{scl}) is only 1.4\%. 
 The wide range of ages ($\sim\!2\!-\!10$ Gyr) and metallicities ($-1\!\lesssim\![Fe/H]\!\lesssim\!0.0$)
 in the Coma cluster dwarfs suggests non-unique
 evolutionary histories for these galaxies \citep[][and references therein]{lisker07,smith08}.
 While some of the oldest red dwarfs, confined to the cluster core, may belong to the primordial
 cluster population \citep{smith08}, others could have transformed from luminous late-type spirals,
 after their infall into the cluster \citep[Fig.~\ref{image}; also see][]{boselli08,barazza09}. 
 The surface brightness profiles of dwarf ($-18\!\leq\!M_B\!\leq\!-16$) galaxies in
 the Coma cluster also show that while the galaxies fitted with a single S{\'e}rsic profile
 (dEs) could come from star-forming dwarfs, the dwarf lenticulars (fitted with
 S{\'e}rsic\,$+$\,exponential profile) are evolved spirals that were harassed during infall
 \citep[][also see Porter et al. 2008]{aguerri05}. 
 
 Figs.~\ref{scl} and \ref{ha-hd} provide further clues to the evolution path adopted by
 dwarfs. While some blue galaxies show ongoing starburst (right-hand
 ridge in Fig.~\ref{ha-hd}), star formation in others, especially those falling into clusters,
 must be suddenly quenched (Figs.~\ref{scl} and \ref{radial}). On the other
 hand, the red galaxies trace the high density regions, suggesting a relatively stronger star
 formation-density relation throughout the supercluster (Fig.~\ref{cmr} \& \ref{scl}). 

 The range of values of various relevant measured parameters, explored in this article for the
 dwarf galaxies in the Coma supercluster do not allow us to sub-divide our sample into galaxies
 with widely different evolutionary histories. Instead, by considering 
 all dwarfs in Coma as a single ensemble, we have been able to provide some
 insight into the complex mix of star formation histories among this vulnerable population.  
 It is interesting to note that contrary to the results presented in this article, \citet{yan09}
 find that the giant k+A galaxies ($0.1\!<\!\rm z\!<\!0.8$) show an environmental dependence similar
 to that of blue galaxies. In a study of $0.05-0.5\,L^*$ galaxies in a cluster at z\,$=$\,0.54,
 \citet{delucia09}
 found only 6 post-starburst galaxies amongst the spectroscopic cluster members, thus
 suggesting that the transformation of star-forming infalling galaxies to the faint red ones
 found in the low redshift clusters must have occurred primarily through
 physical processes that do not lead to a post-starburst phase.  
 In order to establish and understand the mass dependence in the occurrence and nature of
 k+A galaxies, a homogeneous analysis of the spectroscopic properties of galaxies spanning a wide
 range of $M^*$, environments and different epochs is required. We intend to return to this in future.

 \section{Conclusions}
 \label{conclusions}
  
 By using the SDSS DR7 data from the spectroscopic galaxy catalogue for the dwarf ($r\!\leq\!17.77;
 z\!>\!15$) galaxies in the Coma supercluster (z\,$=\!0.023$), we have shown that:
 \begin{itemize}
   \item The mean $(g\!-\!r)$ colour of dwarf galaxies in and around the Coma cluster increases
         from 0.3 at $\sim\!3$R$_{\rm virial}$ to 0.7 in the core. 
   \item In the Coma supercluster, the k+A dwarf galaxies are mostly confined to the 
         cluster infall regions, and occasionally galaxy groups, suggesting that
	 most of the dwarf galaxies falling into the Coma and Abell~1367 clusters are likely to
	 experience sudden quenching of star formation and hence are observable in the k+A phase.
         On the contrary, dwarfs falling into galaxy groups embedded in the filament are likely to
	 suffer slow truncation.
	 \item The blue colour of some k+A dwarfs in the Coma cluster lying within its virial radius, suggests
	 that the
	 star formation in these galaxies was quenched rapidly within the last $500$\,Myr. In contrast, the
	 last major episode of star formation in the red k+A galaxies should have ceased on a timescale
	 of $\sim\!1$\,Gyr. 
   \item Assuming the red dwarf galaxies to be passively evolving, they show a stronger
    star formation-density relation relative to their bluer counterparts.  
   \item A significant fraction of all red galaxies in the Coma supercluster show some absorption in
         H$_\delta$ (0--3\,\AA), suggesting that star formation in them could have been quenched
         within the last Gyr.   
   \item The populations of k+A and red dwarfs increases towards the cluster core in Coma from
         almost twice the virial boundary, while that of the blue ones decreases steadily.
         The k+A galaxy distribution is flatter relative to red dwarfs, and they are not as concentrated
         in the core of the cluster as the red dwarfs, suggesting
         that the k+A galaxies preferentially avoid very dense and underdense regions in equal
         measures.
	       
 \end{itemize}
 
 \section{Acknowledgments}
 
 This research has made use of the SAO/NASA Astrophysics Data System,
 and the NASA/IPAC Extragalactic Database (NED).  
 Funding for the SDSS and SDSS-II has been provided by the Alfred
 P. Sloan Foundation, the Participating Institutions, the National Science
 Foundation, the U.S. Department of Energy, the National Aeronautics and Space
 Administration, the Japanese Monbukagakusho, the Max Planck Society, and the Higher
 Education Funding Council for England. The SDSS Web Site is http://www.sdss.org/.
 SM is supported by grants from ORSAS, UK, and the University of 
 Birmingham. We thank the anonymous referee for constructive
 comments which have greatly helped in improving this article. 
 

 \label{lastpage}

 \end{document}